\documentclass[12pt]{iopart}

\usepackage{amsfonts}
\input epsf
\begin{document}

\title[]{Uses of zeta regularization in QFT with boundary conditions:
a cosmo-topological Casimir effect\footnote{Talk given at the
Seventh International Workshop {\it Quantum Field Theory under the
Influence of External Conditions, QFEXT'05}, Barcelona, September
5-9, 2005.}}

\author{Emilio Elizalde}

\address{Instituto de Ciencias del Espacio (CSIC) \\ \&
Institut d'Estudis Espacials de Catalunya (IEEC/CSIC) \\
Campus UAB, Facultat de Ci\`encies, Torre C5-Parell-2a planta \\
E-08193 Bellaterra (Barcelona) Spain} \ead{elizalde@ieec.uab.es
http://www.ieec.fcr.es/english/recerca/ftc/eli/eli.htm}

\begin{abstract}
Zeta regularization has proven to be a powerful and reliable tool
for the regularization of the vacuum energy density in ideal
situations. With the Hadamard complement, it has been shown to
provide finite (and meaningful) answers too in more involved cases,
as when imposing physical boundary conditions (BCs) in two-- and
higher--dimensional surfaces (being able to mimic, in a very
convenient way, other {\it ad hoc} cut-offs, as non-zero depths).

Recently, these techniques have been used in calculations of the
contribution of the vacuum energy of the quantum fields pervading
the universe to the cosmological constant (cc). Naive counting of
the absolute contributions of the known fields lead to a value which
is off by as much as 120 orders of magnitude, as compared with
observational tests, what is known as the {\it cosmological constant
problem}. This is very difficult to solve and we do not address that
question directly.

What we have considered ---with relative success in several
approaches of different nature--- is the {\it additional}
contribution to the cc coming from the non-trivial topology of space
or from specific boundary conditions imposed on braneworld models
(kind of cosmological Casimir effects). Assuming someone will be
able to prove (some day) that the ground value of the cc is zero, as
many had suspected until very recently, we will then be left with
this incremental value coming from the topology or BCs. We show that
this value can have the correct order of magnitude ---corresponding
to the one coming from the observed acceleration in the expansion of
our universe--- in a number of quite reasonable models involving
small and large compactified scales and/or brane BCs, and
supergravitons.

\end{abstract}

%Uncomment for PACS numbers title message
%\pacs{00.00, 20.00, 42.10}
% Keywords required only for MST, PB, PMB, PM, JOA, JOB?
%\vspace{2pc}
%\noindent{\it Keywords}: Article preparation, IOP journals
% Uncomment for Submitted to journal title message
%\submitto{\JPA}
% Comment out if separate title page not required
\maketitle

\section{Introduction}
As crudely stated by Jaffe \cite{jaffe1}, experimental confirmation
of the Casimir effect does not establish by itself the reality of
zero point fluctuations. He explains this via the example of the
electromagnetic field, where the energy of a smooth charge
distribution, $\rho(x)$, can be precisely calculated from the energy
stored in the electric field, a formula which arguably cannot be
taken as evidence for the electric field itself being real.
Fortunately, propagating electromagnetic waves are detected all the
time. The moral: in the case of the Casimir forces one should look
for {\it direct} evidence of vacuum fluctuations. Have they been
found yet? As of today, the answer is very controversial.\footnote{I
could check that personally, when I delivered this lecture at the
Workshop.} Since GR has much wider consensus, I here propose a
search at the cosmological level. In fact, almost everybody admits
that any sort of energy will always gravitate. Thus, the energy
density of the vacuum, more precisely, the vacuum expectation value
of the stress-energy tensor,
\begin{eqnarray} \langle T_{\mu \nu} \rangle \equiv - {\cal E}
g_{\mu \nu},
\end{eqnarray}
appears on the rhs of Einstein's equations
\begin{eqnarray} R_{\mu\nu}-\frac{1}{2}g_{\mu\nu}R=-8\pi
G(\tilde{T}_{\mu\nu}-{\cal E}g_{\mu\nu}). \end{eqnarray} It
therefore affects {\it cosmology}: there is a contribution  $\tilde
T_{\mu \nu}$ of excitations above the vacuum, equivalent to a {\it
cosmological constant}  $\lambda =8\pi G{\cal E}$. Recent data yield
\cite{sdsscol1}
\begin{eqnarray}
\lambda = (2.14\pm 0.13 \times 10^{-3}\ \hbox{eV})^{4} \ \sim \ 4.32
\times 10^{-9}\ \hbox{erg/cm}^3. \label{j1}
\end{eqnarray}
At issue is then the belief that zero point fluctuations will
contribute in an essential way to the cosmological constant (cc),
e.g., they will be of the same order of magnitude.

Different rigorous techniques have been used recently in order to
perform this calculation, the result being that the absolute
contributions of the known quantum fields (all of which couple to
gravity) lead to a value which is off by roughly  120 orders of
magnitude ---kind of a modern (and indeed very thick!) ether.
Extremely severe cancelations should occur. Observational tests, as
advanced, see nothing (or very little) of it, what leads to the
so-called cosmological constant problem \cite{ccp1}. This problem is
at present very difficult to solve and we will here {\it not}
address such hard question directly. Some {\it almost} successful
attempts at solving the problem deserve to be mentioned, as the
clever approaches by Baum and Hawking, and Polchinski's phase
ambiguity found in Coleman's solution \cite{ccp2}.

What we do consider here ---with relative success in quite different
approaches--- is the {\it  additional} contribution to the cc coming
from the {\it non-trivial topology} of space or from specific {\it
boundary conditions} imposed on braneworld and other models. This
can be viewed as kind of a Casimir effect at cosmological scale: a
{\it cosmo-topological Casimir effect}. Assuming someone will be
able to prove (some day) that the ground value of the cc is {\it
zero} (as many had suspected until very recently),\footnote{What
would, by the way, correspond to the convention of normal ordering
in QFT in ordinary, Euclidean backgrounds.} we will be left with
this incremental value coming from the topology or BCs. We show here
that this value has the correct order of magnitude, e.g., the one
coming from the observed acceleration in the expansion of our
universe, in three different types of models, involving: (a) small
and large compactified scales, (b) dS and AdS worldbranes, and (c)
supergravitons.

\section{Simple model with large and small dimensions}
Consider a universe with a space-time such as: $\mathbf{R^{d+1}}
\times \mathbf{T}^p\times \mathbf{T}^q$, $\mathbf{R^{d+1}} \times
\mathbf{T}^p\times \mathbf{S}^q, \ldots$, which are very simple
models for the space-time topology. A free scalar field pervading
the universe will satisfy $(-\Box +M^2) \phi =0$, restricted by the
appropriate boundary conditions (e.g., periodic, in the first case).
Here, $d\geq 0$ stands for a possible number of non-compactified
dimensions. Recall that the physical contribution to the vacuum or
zero-point energy $< 0 | H |  0 >$ ($H$ is the Hamiltonian  and $|
0 >$ the vacuum state) is obtained after subtracting $E_C = \left.
< 0 | H | 0
>\right|_R  - \left. < 0 | H | 0
>\right|_{R\rightarrow \infty}$
($R$ being a compactification length), what gives rise to the finite
value of the Casimir energy $E_C$, which will depend on $R$, after a
regularization/renormalization procedure is carried out. We discuss
the Casimir energy {\it density} $\rho_C=E_C/V$, for either a finite
or an infinite volume of the spatial section of the
universe.\footnote{From now on we assume that all diagonalizations
already correspond to energy densities, and the volume factors will
be replaced at the end.} In terms of the spectrum: $< 0 | H |  0 > =
\frac{1}{2} \sum_n \lambda_n$,  the sum over $n$ involving, in
general, several continuum and several discrete indices.

The physical vacuum energy density corresponding to the contribution
of a scalar field, $\phi$ in a (partly) compactified spatial section
of the universe is\footnote{Note that this is just the contribution
to $\rho_V$ coming from this field; there might be other, in
general.}
\begin{eqnarray}
 \rho_\phi =\frac{1}{2}
\sum_{\mbox{\bf k}} \frac{1}{\mu} \left(k^2 +M^2\right)^{1/2},
\label{c2} \end{eqnarray} where  $\mu$ is the usual mass-dimensional
parameter to render the eigenvalues dimensionless (we take $\hbar =c
=1$ but will insert the dimensionfull units  at the end). The mass
$M$ of the field will be kept different from zero (a tiny mass can
never be excluded) and its allowed value will be constrained later.
A lack of this simplified model: the coupling of the scalar field to
gravity should be considered (see, e.g., \cite{pr1} and the
references therein). However, taking it into account does not change
the order of magnitude of the results. The renormalization of the
model is rendered much more involved, and one must enter a
discussion on the orders of magnitude of the different
contributions, which yields, in the end, an ordinary perturbative
expansion, the coupling constant being finally re-absorbed into the
mass of the scalar field. Owing, essentially, to the smallness of
the resulting mass for the scalar field, one can prove that,
quantitatively, the difference in the final result is of some
percent only. Another consideration: our model is stationary, while
the universe is expanding. Again, this effect can be dismissed at
the level of our order-of-magnitude calculation, since this
contribution is clearly less than the one we will get ---taken the
present value of the expansion rate $\Delta R /R \sim 10^{-10}$ per
year, or from direct consideration of the Hubble coefficient. In any
case, these refinements are left for future work. Here, to focus
just on  the essential issue, we perform a static calculation and
the value of the Casimir energy density and cc to be obtained will
correspond to the present epoch. They are bound to change with time.

\subsection{Regularization of the vacuum energy density}
For a ($p,q$)-toroidal universe, with $p$ the number of large and
$q$ of small dimensions: \begin{eqnarray} \rho_\phi =
\frac{1}{a^pb^q} \sum_{\mbox{\bf n}_p, \mbox{\bf m}_q=-\mathbf{
\infty}}^{\mathbf{\infty}} \left( \frac{1}{a^2}\sum_{j=1}^p n_j^2 +
\frac{1}{b^2} \sum_{h=1}^q m_h^2 +M^2 \right)^{(d+1)/2+1},\label{t2}
\end{eqnarray} which corresponds to
all large (resp. all small) compactification scales being the same.
The squared mass of the field should be divided
by $4\pi^2\mu^2$, but we have renamed it again $M^2$ to simplify.
 We also dismiss the mass-dim factor $\mu$, easy to recover later.

For a ($p$-toroidal, $q$-spherical)-universe,
\begin{eqnarray} \rho_\phi =\frac{1}{a^pb^q} \sum_{\mbox{\bf
n}_p=-\mathbf{\infty}}^{\mathbf{\infty}} \sum_{l=1}^\infty
P_{q-1}(l)  \left( \frac{4\pi^2}{a^2}\sum_{j=1}^p n_j^2 +
 \frac{l(l+q)}{b^2}  +M^2 \right)^{(d+1)/2+1},\label{ts2}
\end{eqnarray}  $P_{q-1}(l)$ being a polynomial in $l$
of degree $q-1$. We assume that $d=3-p$ is the number of
non-compactified, large spatial dimensions, and $\rho_\phi$ needs to
be regularized. We use the zeta function \cite{zb123}, taking
advantage of our expressions in \cite{jpa2001,eecmp1}. No further
subtraction or renormalization is needed (the subtraction at
infinity is zero, and not even a finite renormalization shows up).
Using the mentioned formulas, that generalize the
 Chowla-Selberg expression to encompass Eqs. (\ref{t2}) and (\ref{ts2}),
we can provide arbitrarily accurate results (even for
different values of the compactification radii \cite{eejmp12}).

For the first case, Eq. (\ref{t2}), we obtain
 \begin{eqnarray}
\hspace*{-30mm} \rho_\phi = -\frac{1}{a^pb^{q+1}} \sum_{h=0}^p
\left(_{\,\displaystyle h\,}^{\,\displaystyle p\,}\right) 2^h
\hspace*{-2mm} \sum_{\mbox{\bf n}_h=1}^{\mathbf{\infty}}
\sum_{\mbox{\bf m}_q=-\mathbf{\infty}}^{\mathbf{\infty}}
\hspace*{-1mm} \sqrt{\frac{\sum_{k=1}^q m_k^2+M^2}{\sum_{j=1}^h
n_j^2 }}  K_1 \left[ \frac{2\pi a}{b} \sqrt{\sum_{j=1}^h n_j^2
\left(\sum_{k=1}^q m_k^2+M^2\right)}\right]. \ \ \ \label{c11}
\end{eqnarray} Now, from the behaviour of the function $K_\nu (z)$
for small values of its argument, $ K_\nu (z) \sim \frac{1}{2}
\Gamma (\nu) (z/2)^{-\nu}, \, z \to 0$, we get, in the case when $M$
is  small, \begin{eqnarray}  && \hspace*{-23mm} \rho_\phi =
-\frac{1}{a^pb^{q+1}} \left\{ M\, K_1 \left( \frac{2\pi a}{b} M
\right)+ \sum_{h=0}^p \left(_{ \, \displaystyle h\,}^{\,
\displaystyle p\,}\right) 2^h \sum_{\mbox{\bf
n}_h=1}^{\mathbf{\infty}} \frac{M}{\sqrt{\sum_{j=1}^h n_j^2 }} \
K_1\left( \frac{2\pi a}{b} M \sqrt{\sum_{j=1}^h n_j^2}
\right)\right.\nonumber \\ && \hspace*{40mm} + \left. {\cal O}
\left[ q\sqrt{1+M^2} K_1\left( \frac{2\pi a}{b}\sqrt{1+M^2}\right)
\right]\right\}. \label{ff0}\end{eqnarray} The only presence of the
mass-dim parameter $\mu$ is as $M/\mu$ everywhere, and this does not
affect the small-$M$ limit, $M/\mu << b/a$. Inserting back  the
$\hbar$ and $c$ factors, we  get
\begin{eqnarray} \rho_\phi = -\frac{\hbar c}{2\pi a^{p+1}b^q}
\left[1+\sum_{h=0}^p \left(_{\,\displaystyle h\,}^{\,\displaystyle
p\,}\right) 2^h \alpha \right]+ {\cal O} \left[ q K_1\left(
\frac{2\pi a}{b}\right) \right], \label{ff1}
\end{eqnarray} where $\alpha$ is a computable finite constant,
obtained as an explicit geometrical sum in the limit $M\rightarrow
0$. It is remarkable that we do get a well defined limit,
independent of $M^2$, provided $M^2$ is small enough.\footnote{Indeed,
 a  physically nice situation turns out to correspond to the
mathematically rigorous case.}

\subsection{Numerical results}
For the most common cases, the constant $\alpha$ in (\ref{ff1}) has
been calculated to be of order $10^2$, and the whole factor, in
brackets, of order $10^7$. This clearly shows the value of a precise
calculation, as the one undertaken here, together with the fact that
just a naive consideration of the dependencies of $\rho_\phi$ on the
powers of the compactification radii, $a$ and $b$, is actually {\it
not enough} in order to get the correct result. Note, moreover, the
non-trivial change in the power dependencies on going from
Eq.~(\ref{ff0}) to Eq.~(\ref{ff1}).

Naturally enough, for the compactification radii at small scales,
$b$, we take the Planck length, $b \sim l_{P(lanck)}$, and for the
large scales, $a$, the present size of the universe, $a\sim R_U$.
With these choices, the order of $a/b$ in the argument of $K_1$ is
as big as: $a/b \sim 10^{60}$.\footnote{Note that the square of this
value yields the 120 orders of magnitude of the QFT cc.}
   The final
expression for the vacuum energy density is independent of the mass
$M$ of the field, provided this is  small enough (eventually zero).
In fact, the last term in Eq. (\ref{ff1}) is exponentially vanishing
(zero, for {\it app}). In ordinary units the bound on the mass of
the scalar field is   $M \leq 1.2 \times 10^{-32}$ eV (e.g.,
physically zero, since it is less by several orders of magnitude
than any bound coming from  SUSY theories). \footnote{Where in fact
scalar fields with low masses of the order of that of the lightest
neutrino do show up \cite{shs1}, which may have observable
implications.}

\begin{table}[bht]

\begin{center}

\begin{tabular}{|c||c|c|c|c|}
\hline \hline $\rho_\phi$ & $p=0$ & $p=1$ & $p=2$ & $p=3$ \\
 \hline \hline
$b=l_P$ & $10^{-13}$ & $10^{-6}$ & 1 & $10^5$ \\ \hline $b=10\, l_P$
&$10^{-14}$ & [$10^{-8}$] & $10^{-3}$ & $10$
 \\ \hline
$b=10^2 l_P$  &$10^{-15}$ & ($10^{-10}$) & $10^{-6}$ & $10^{-3}$
 \\ \hline
$b=10^3 l_P$ & $10^{-16}$ & $10^{-12}$ & [$10^{-9}$] & ($10^{-7}$)
\\ \hline $b=10^4 l_P$ & $10^{-17}$ & $10^{-14}$ & $10^{-12}$ &
$10^{-11}$ \\ \hline $b=10^5 l_P$ & $10^{-18}$ & $10^{-16}$ &
$10^{-15}$  & $10^{-15}$
 \\ \hline \hline \end{tabular}

\caption{{\protect\small Vacuum energy density contribution (orders
of magnitude, omitting the minus sign everywhere), in units of
erg/cm$^3$, Eq.~(\ref{j1}). In brackets, the values that more
exactly match the one for the cosmological constant coming from
observations, and in parenthesis the otherwise closest
approximations.}}
\end{center}
\end{table}

\vspace*{-4mm}

By replacing such values we obtain Table 1. The total number of
large space dimensions is three (our universe). Good coincidence in
absolute value with the observational value is obtained for  $p$
large and $q=p+1$ small compactified dimensions, $p=0,\ldots,3$, and
this for the small compactification length, $b$, of the order of 10
to 1000 times the Planck length $l_P$ (a most reasonable range,
according to string theory).  The $p$ large and $q$ small dimensions
are {\it not} all that are supposed to exist: $p$ and $q$ refer to
the {\it compactified} ones only. There may be non-compactifed
dimensions, what translates into a modification of the formulas
above,  but does not change the {\it order of magnitude} of the
final numbers (see e.g.~\cite{zb123} for an elaboration on this
technical point). Finally, simple power counting is {\it unable} to
provide the correct order of magnitude of the results here obtained.
One should observe however that the sign of the cc is a problem with
these oversimplified models (generically they get it wrong!). This
is no longer so with the more elaborate theories involving bosons
and fermions to be considered below where, using quite natural
boundary conditions, an expanding universe can be obtained.

\section{Braneworld models}
Braneworld theories may help to solve both the hierarchy problem and
the cc problem. The bulk Casimir effect can play an important role
in the construction (radion stabilization) of braneworlds. We have
calculated the bulk Casimir effect (effective potential)  for
conformal and for  massive scalar fields \cite{enoo2003}. The bulk
is a 5-dim AdS or dS space, with 2 (or 1) 4-dim dS branes (our
universe). The results obtained are consistent with observational
data. We present a summary of those results here.

For the case of two dS$_4$ branes (at $L$ separation) in a dS$_5$
background (it becomes a one-brane configuration as $L\to \infty$)
the Casimir energy density and effective potential, for a
conformally invariant scalar-gravitational theory $ {\cal S} =
{1\over 2}\int d^{5}x \sqrt{g} \left[ -g^{\mu\nu}\partial_{\mu} \phi
\partial_{\nu} \phi + \xi_{5}R^{(5)} \phi^2 \right] $,
$\xi_{5}=-3/16$, with $R^{(5)}$ the curvature and $ds^2 =
g_{\mu\nu}dx^{\mu}dx^{\nu} ={\alpha^2 \over \sinh^2 z} \left( dz^2 +
d\Omega _{4}^2 \right)$ the Euclidean metric of the 5-dim AdS bulk,
$ d\Omega_{4}^2 = d\xi^2 + \sin^2 \xi d\Omega_3^2$ ---for the 4-dim
manifold, $M_4$, with $\alpha$ the AdS radius, related to the cc of
the AdS bulk, and $d\Omega_3$ the metric on the 3-sphere, of radius
$\cal R$--- are obtained as follows. For the one-brane Casimir
energy density (pressure), we get
\begin{eqnarray}
\hspace*{-14mm} {\cal E}_{\mbox{Cas}} = \frac{\hbar c}{2L {\rm
Vol}(M_{4})}\zeta \left( -{1\over 2}| L_5 \right) = - \frac{\hbar c
\pi^3 }{36 L^6} \left[ \frac{\pi^2}{315} - \frac{1}{240}
\left(\frac{L}{\cal R} \right)^2 + {\cal O} \left(\frac{L}{\cal R}
\right)^4\right].
\end{eqnarray}
%which is about  ten times larger than the ordinary Casimir effect:
%${\cal E}_{\mbox{\small CE}} = -\frac{\hbar c \pi^2}{240 \, L^4}$
%(about 100 dynes/cm$^2$ at 100 nm).
For the one-loop effective potential, we have
\begin{eqnarray}
V={1\over 2L {\rm Vol}(M_{4})}\log \det (L_5/ \mu^2),
\end{eqnarray}
where $ L_5=-\partial _{z}^2 -\Delta ^{(4)} -\xi_{5}R^{(4)}=L_1+L_4,
$ and $\log \det L_5 =\sum_{n,\alpha} \log (\lambda^2_{n} +
\lambda^2_{\alpha}) = -\zeta' (0|L_5)$. In the one-brane limit  $L
\rightarrow \infty$, $ \zeta' (0|L_5)= {1\over 3 {\cal R}} \left[
\zeta_{H}\left( -4 , {3\over 2} \right)-{1\over 4} \zeta_{H}\left(
-2 , {3\over 2} \right) \right]  = 0$. And the small distance
expansion for the effective potential yields (up to an overall
factor)
\begin{eqnarray}
&& \hspace*{-22mm} \zeta' (0|L_5) = \frac{\zeta' (-4)}{6}\,
\frac{\pi^4 {\cal R}^4}{L^4} +  \frac{\zeta' (-2)}{12}\, \frac{\pi^2
{\cal R}^2}{L^2} + \frac{1}{24} \left[  \zeta_H' (-4,3/2)
 - \frac{1}{2} \zeta_H' (-2,3/2)\right]
\ln \frac{\pi^2 {\cal R}^2}{L^2}  \nonumber \\ &&  + \frac{\zeta'
(0)}{6}
 \left[  \zeta_H' (-4,3/2) - \frac{1}{2} \zeta_H' (-2,3/2)\right] +
\frac{1}{24} \zeta_H' (-4,3/2)  \nonumber \\ &&  + \frac{1}{36}
\left[ \frac{1}{8} \zeta_H' (-4,3/2) - \frac{1}{3} \zeta_H'
(-6,3/2)\right] \frac{L^2}{{\cal R}^2} + {\cal O} \left( \frac{L^4}{
\pi^4 {\cal R}^4}\right)   \\ && \hspace*{-20mm}  \simeq 0.129652 \,
\frac{{\cal R}^4}{L^4} - 0.025039 \, \frac{{\cal R}^2}{L^2} -
0.002951 \, \ln \frac{{\cal R}^2}{L^2}  -0.017956 - 0.000315
\frac{L^2}{{\cal R}^2} + \cdots \nonumber
\end{eqnarray}

On the other hand, the effective potential for the massive scalar
field model is obtained to
be \begin{eqnarray}
V&=&{1\over 2L {\rm Vol}(M_{4})}\log \det (L_5/ \mu^2),  \\
L_5&\equiv & -\partial _{z}^2 +m^2 l^2 \sinh ^{-2}z -\Delta ^{(4)}
 - \xi_{5}R^{(4)} = L_1 +L_4  \ \ \  (AdS),\nonumber \\
L_5&\equiv & -\partial _{z}^2 +m^2 \cosh ^{-2}z -\Delta ^{(4)}
 - \xi_{5}R^{(4)} = L_1 +L_4  \ \ \ \ \ \ (dS). \nonumber
\end{eqnarray}
For the small mass limit (with $L$ not large), it yields
\begin{eqnarray}
\hspace*{-12mm}  \zeta' (0|L_5)&\simeq& \frac{a\rho +a^2\rho^2}{48}
-\frac{\pi^2}{144} \left\{\frac{a \rho^2}{2} + \left[
2\zeta'(-4,3/2)-\zeta'(-2,3/2)\right] \rho\right\}\nonumber
\\ && -\frac{\pi^4}{4370} \left[ 2\zeta'(-4,3/2)-
\zeta'(-2,3/2)\right] \rho^2 +{\mathcal O} (m^6),  \\ &&
\hspace*{25mm}  a\equiv {\pi^2{\mathcal R}^2 \over L^2}, \quad \rho
\equiv {m^2l^2 \over \pi^2}\, {\tanh ( L/2l) \over L/2l},\nonumber
\end{eqnarray} while for the large mass limit (with $L$ not small),
it is
\begin{eqnarray}
  \zeta' (0|L_5) &=& - \frac{4 m^2l^3}{3\mathcal R} \,
{\arctan (\sinh  L/2l) \over \sinh  ( L/2l)} + \cdots\,,
\end{eqnarray} which is now non-zero (unlike in previous
calculations, which turned a vanishing value) and can fit the
observed order of magnitude under appropriate conditions.

\section{Supergraviton theories}
Finally, we have also computed the effective potential for some
multi-graviton models with supersymmetry \cite{ss1}. In one case,
the bulk is a flat manifold with the torus topology {\bf R} $\times$
{\bf T}$^3$, and it can be shown that the induced cosmological
constant can  be rendered {\it positive} due to topological
contributions \cite{cezplb05}. Previously, the case of {\bf R}$^4$
had been considered. In the multi-graviton model the induced
cosmological constant can indeed be positive, but only if the number
of massive gravitons is sufficiently large, what is not easy to fit
in a natural way. In the supersymmetric case, however, the
cosmological constant turns out to be positive just by imposing
 anti-periodic BC in the fermionic sector. An essential issue in our
model is to allow for non-nearest-neighbor couplings.

The multi-graviton model is defined by taking $N-$copies of the
fields with graviton $h_{n\mu\nu}$ and St{\"u}ckelberg fields
$A_{n\mu}$ and $\varphi_n$. Our theory is defined by a Lagrangian
which is a generalization of the one in \cite{KS}. It reads
\begin{eqnarray} \label{KS9} && \hspace*{-16mm} {\cal L}=
\sum_{n=0}^{N-1}\left[ -{1 \over 2}\partial_\lambda h_{n\mu\nu}
\partial^\lambda h_n^{\mu\nu}
+ \partial_\lambda h^\lambda_{n\mu}\partial_\nu h_n^{\mu\nu}
 - \partial_\mu h_n^{\mu\nu}\partial_\nu h_n
+ {1 \over 2}\partial_\lambda h_n\partial^\lambda h_n \right.
\nonumber\\
&& - {1 \over 2}\left(m^2\Delta h_{n\mu\nu}\Delta h_n^{\mu\nu} -
\left(\Delta h_n\right)^2 \right) - 2 \left(m\Delta^\dagger A_n^\mu
+
\partial^\mu \varphi_n\right) \left(\partial^\nu h_{n\mu\nu} -
\partial_\mu h_n\right)
\nonumber\\
&& \left. - {1 \over 2}\left(\partial_\mu A_{n\nu} - \partial_\nu
A_{n\mu}\right) \left(\partial^\mu A_n^\nu - \partial^\nu
A_n^\mu\right) \right]\ . \end{eqnarray}
 The $\Delta$ and $\Delta^\dagger$ are
difference operators, which operate on the indices $n$ as
$\Delta\phi_n \equiv \sum_{k=0}^{N-1}a_k \phi_{n+k}$,
$\Delta^\dagger\phi_n \equiv \sum_{k=0}^{N-1}a_k \phi_{n-k}$,
$\sum_{k=0}^{N-1}a_k = 0$, where the $a_k$ are $N$ constants and the
$N$ variables $\phi_n$ can be identified with periodic fields on a
lattice with $N$ sites if the  periodic boundary conditions,
$\phi_{n+N}=\phi_n$, are imposed. The latter condition  assures that
$\Delta$ becomes the usual differentiation operator in a properly
defined continuum limit.

In the case when anti-periodic boundary conditions are imposed in
the fermionic sector, the situation changes completely with respect
to the bosonic one, since the fermionic mass spectrum becomes quite
different. The one-loop effective potential in the anti-periodic
case is calculated to be
\begin{eqnarray} && \hspace*{-13mm} V_{eff}=\frac{M^4_1}{4\pi^2}
\left(\ln\frac{M_1^2}{\mu_R^2}-\frac{3}{2}\right)
-\frac{4M_1^{4}}{3\pi^2}\, \int_{1}^{\infty}\,du\;G(M_1
ru)\,(u^2-1)^{3/2} \nonumber\\&&\qquad -\frac{\tilde M^4_0}{4\pi^2}
\left(\ln\frac{\tilde M_0^2}{\mu_R^2}-\frac{3}{2}\right)
+\frac{4\tilde M_0^{4}}{3\pi^2}\, \int_{1}^{\infty}\,du\;G(\tilde
M_0 ru)\,(u^2-1)^{3/2} \nonumber\\&&\qquad-\frac{\tilde
M^4_1}{8\pi^2} \left(\ln\frac{\tilde
M_1^2}{\mu_R^2}-\frac{3}{2}\right) +\frac{2\tilde M_1^{4}}{3\pi^2}\,
\int_{1}^{\infty}\,du\;G(\tilde M_1 ru)\,(u^2-1)^{3/2}
\nonumber\\&=& -\frac{m^4}{36\pi^2}\,\log\frac{2^{16}}{3^9}+V_T,
\label{veff2}\end{eqnarray} where $V_T$ is the sum of all the
topological contributions. Note that the first term on the rhs is
always negative, but the whole effective potential can be positive,
due to the presence of the topological term. Thus, in the regime
$mr\ll 1$ one has \begin{eqnarray} \hspace*{-8mm}
V_T\sim\frac{1}{8\pi^2r^4}\quad\Longrightarrow\quad
V_{eff}>0\quad\mbox{for}\quad
mr<\left(\frac{2}{9}\,\log\frac{2^{16}}{3^9}\right)^{-1/4}\sim 1.4,
 \end{eqnarray} while in the opposite regime, $mr\gg 1$, we can see that
 the topological contribution (although
still positive) is negligible, and the effective potential remains
negative. In Fig.~1, the corresponding plot of the full effective
potential, Eq. (\ref{veff2}), is depicted as a function of $y\equiv
mr$. The change of sign in the correct region is clearly observed.
\begin{figure}[htb]
\vskip-8mm \centerline{\epsfxsize=11.5cm \epsfbox{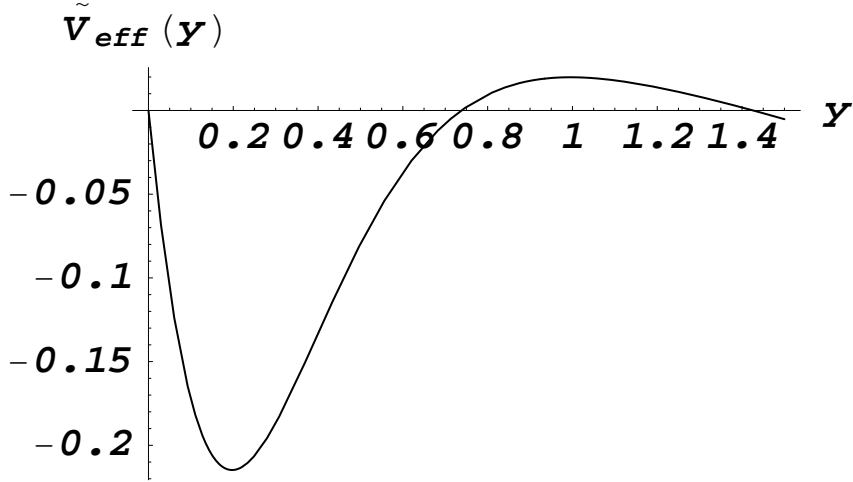}}
\vskip-2mm \caption{{\protect\small Plot of
$\tilde{V}_{eff}(y)\equiv r^4V_{eff}(r)$, Eq. (\ref{veff2}), as a
function of $y\equiv mr$. }} \label{f3}
\end{figure}

To summarize, in the case of the torus topology we have obtained
that the topological contributions to the effective potential have
always a fixed sign, which depends on the BC one imposes.  They are
negative for periodic fields, and positive for anti-periodic fields.
But topology provides then a mechanism which, in a most natural way,
permits to have a positive cc in the multi-supergravity model with
anti-periodic fermions. Moreover, the value of the cc is regulated
by the corresponding size of the torus. We can most naturally use
the minimum number, $N = 3$, of copies of bosons and fermions, and
show that ---as in the first, much more simple example, but now with
the right sign!--- within our model the observational values for the
cosmological constant, Eq.~(\ref{j1}), can be matched, by making
very reasonable adjustments of the parameters involved. As a
byproduct, the results that we have obtained \cite{cezplb05} may
also be relevant in the study of electroweak symmetry breaking in
models with similar type of couplings, for the deconstruction issue.
\medskip

\noindent{\bf Acknowledgments.} Based on work done in part in
collaboration with G. Cognola, S. Nojiri, S.D. Odintsov, S. Ogushi
and S. Zerbini. The financial support of DGICYT BFM2003-00620, SEEU
PR2004-0126, and CIRIT 2001SGR-00427 is gratefully acknowledged.

\end{document}